\documentclass[cits]{PoS}

\title{mm-VLBI observations of the active galaxy 3C\,111 in outburst}

\ShortTitle{mm-VLBI observations of the active galaxy 3C\,111 in outburst}

\author{\speaker{Robert Schulz}$^{a,b}$, Matthias Kadler$^a$, Eduardo Ros$^{c,d}$, Thomas P. Krichbaum$^d$, Christoph Grossberger$^{b,a}$, Cornelia M\"uller$^{b,a}$, Karl Mannheim$^a$, Iv\'{a}n Agudo$^{e,f}$, Hugh D. Aller{$^g$}, Margo F. Aller{$^g$}\\ 
        \llap{$^a$}Lehrstuhl f\"{u}r Astronomie, Universit\"{a}t W\"{u}rzburg, 
        	Campus Hubland Nord, Emil-Fischer-Strasse 31, 97074 W\"{u}rzburg, Germany\\
        \llap{$^b$}Dr. Remeis Sternwarte \& ECAP, Universit\"{a}t Erlangen-N\"{u}rnberg, 
            Sternwartstr. 7, 96049 Bamberg, Germany\\
        \llap{$^c$}Departament d'Astronomia i Astrof\'{\i}sica, Universitat de Val\`{e}ncia, 
            46100 Burjassot, Val\`{e}ncia, Spain\\
        \llap{$^d$}Max-Planck-Institut f\"{u}r Radioastronomie, 
        	Auf dem H\"{u}gel 69, 53121 Bonn, Germany\\
        \llap{$^e$}Instituto de Astrof\'{i}sica de Andaluc\'{i}a, CSIC, 
        	Apartado 3004, 18080, Granada, Spain\\
        \llap{$^f$}Institute for Astrophysical Research, Boston University, 
        	725 Commonwealth Avenue, Boston, MA 02215, USA\\
        \llap{$^g$}Department of Astronomy, University of Michigan, 
        	817 Dennison Building, Ann Arbor, MI 48109-1042, USA\\

        E-mail: \email{robert.schulz@physik.uni-wuerzburg.de} 
}

\abstract{The broad-line radio galaxy 3C\,111 exhibited a major flux density outburst in 2007. Here, we present imaging and preliminary kinematic results of the jet, based on three millimetre-VLBI observations (86\,GHz) using the Global Millimeter VLBI Array (GMVA) covering one year just after the radio flare. The GMVA data allow us to study this outburst with unprecedented image fidelity at highest (sub-parsec) resolution. On these scales, the outburst is resolved into a complex series of plasma components forming an intriguing bent structure. Within 1\,mas from the jet base, ejections vary in position angle and components move with an apparent velocity of $\sim$3.7\,c, significantly slower than the maximum velocity observed with cm-VLBI on scales beyond 1\,mas.}

\FullConference{11th European VLBI Network Symposium \& Users Meeting,\\
		October 9-12, 2012\\
		Bordeaux, France}

\begin{document}

\section{Introduction}

The active galaxy 3C\,111 is a broad-line radio galaxy with a classic FRII morphology on kpc-scales \cite{Fanaroff1974,Linfield1984} at a redshift of $z=0.049$. It has been extensively studied at all bands of the electromagnetic spectrum (e.g. \cite{Chatterjee2011,Hartman2008,Kadler2008} and references therein). Long-term radio lightcurves show 3C\,111 to be a strong and variable source (e.g., \cite{Trippe2011}) with a major outburst above 10\,Jy at 90\,GHz in 1996 \cite{Alef1998}. This led to the ejection of a new feature into the pc-scale jet, which shaped the jet morphology over ten years as seen on 15\,GHz VLBA images \cite{Kadler2008}.  At 15\,GHz and 43\,GHz apparent velocities of up to 6\,c were observed \cite{Chatterjee2011,Kadler2008}. The angle of the pc-scale jet to the line of sight at 15\,GHz and 43\,GHz is rather small for a radio galaxy with values of $\sim$15$-$20$^\circ$ \cite{Cohen2007,Jorstad2005,Kadler2008}.

\section{Observation}
In 2007, a major outburst comparable in strength to the one in 1996 occurred \cite{Trippe2011}. We performed three successive observations on 2007-10-13, 2008-05-11 and 2008-10-14 with the Global Millimeter VLBI Array (GMVA\footnote{http://www.mpifr-bonn.mpg.de/div/vlbi/globalmm/}) over a period of one year to study the jet evolution in response to this event at the highest angular resolution. 
Table~\ref{tab:GMVA} gives an overview of the characteristics of the GMVA antennas involved at the time of our observations, which included most of the available mm-wavelength radio telescopes in Europe and the USA. With a maximum baseline length of about 11000\,km, we achieved a resolution of $\sim$45\,$\mu$as, which corresponds to a projected linear scale\footnote{Assuming H$_0$ = 71\,km\,s$^{-1}$\,Mpc$^{-1}$, $\Omega_\Lambda$ = 0.73 and $\Omega_m$ = 0.27 (1\,mas = 0.95\,pc; 1\,mas\,yr$^{-1}$ = 3.1\,c)} of $\sim$0.043\,pc for 3C\,111.
\par
The correlation of the GMVA data was performed with the VLBI correlator at the MPIfR (Bonn, Germany). The Astronomical Imaging Processing System (\textsc{aips}) software package \cite{Greisen2003} was used for the a-priori calibration. The delays and fringe rates of the interferometric phases were corrected with the fringe-fitting algorithm. This step is especially important for mm-VLBI, because of the decreasing coherence length of the atmosphere with higher frequencies (see \cite{Marti-Vidal2012} for more details). The measured system temperatures and antenna gains were applied for amplitude calibration. Hybrid imaging and model-fitting was done in \textsc{difmap} \cite{Shepherd1994}. 

\begin{table}
	\centering
	\begin{tabular}{c|c|c|c|c|c}
		\hline
		Station & Diameter & SEFD & Epoch 1 & Epoch 2 & Epoch 3	\\ 
				&   [m]    & [Jy] & 2007-10-13 & 2008-05-11 & 2008-10-14 \\
		\hline
		\hline
		Mets\"{a}hovi & 14 & 17647 & \checkmark & \checkmark$^2$ & \textsf{X}\\
		Onsala & 20 & 6122 & \checkmark & \checkmark & \checkmark\\
		Effelsberg & 100 & 929 & \checkmark & \checkmark & \checkmark \\
		Plateau de Bure & 35$^*$ & 409 & \checkmark & \checkmark$^2$ & \textsf{X}\\
		Pico Veleta & 30 & 643 & \checkmark$^1$ & \checkmark & \checkmark$^3$\\
		8$\times$VLBA$^\#$ & 25 & 2941 & \checkmark & \checkmark & \checkmark$^4$\\
		\hline
	\end{tabular}
	\caption[]{List of the GMVA stations in 2007 and 2008 with diameter, system equivalent flux density (SEFD) and participation in our observations.\\ 
	{}$^*$Equivalent diameter of a phased interferometric array with 6$\times$15\,m telescopes.\\
	{}$^\#$The eight VLBA stations are North Liberty, Fort Davis, Los Alamos, Pie Town, Kit Peak, Owens Valley, Brewster, Mauna Kea.\\
	{}$^1$No fringes after correlation,{}$^2$Flagged during a-priori calibration,{}$^3$Data were lost due to bad weather, $^4$Data of North Liberty and Fort Davis were flagged during hybrid imaging.}
	\label{tab:GMVA}
\end{table}	


\begin{figure}
\centering
\includegraphics[width=.35\textwidth]{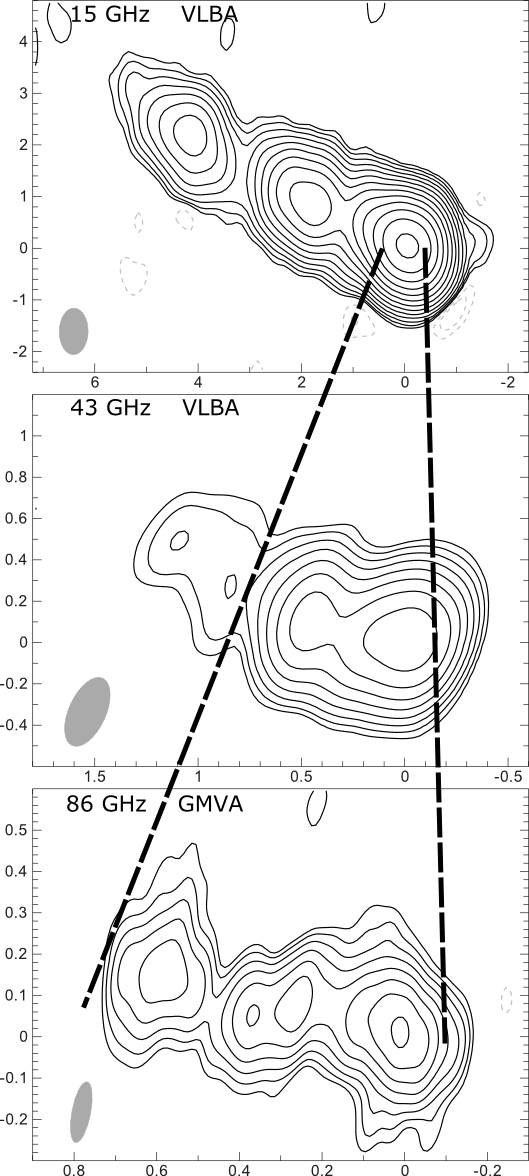}
\caption{Zoom into 3C\,111. Top: 15\,GHz VLBA image (2007-09-06) with a synthesized beam of 0.90$\times$0.56\,mas$^2$ (position angle $-$1$^\circ$) and a peak flux density of 3.23\,Jy/beam; Middle: 43\,GHz VLBA image (2007-09-29) with a synthesized beam of 0.36$\times$0.18\,mas$^2$ (p.a. $-$25$^\circ$) and a peak flux density of 3.36\,Jy/beam; Bottom: 86\,GHz GMVA image (2007-10-13) with a synthesized beam of 0.15$\times$0.045\,mas$^2$ (p.a. $-$11$^\circ$) and a peak flux density of 1.50\,Jy/beam. The lowest contour lines are set to 5$\sigma$ of the rms noise for the first, second and third image respectively, increasing logarithmically by factors of two. The axes show distances relative to the position of the peak emission in units of mas.}
\label{fig:Zoom}
\end{figure}

\section{Results} 

Figure \ref{fig:Zoom} shows images of 3C\,111 at 15, 43 and 86\,GHz, demonstrating the improved angular resolution in going from cm-VLBA to mm-GMVA observations. The two top panels display the straight jet morphology of 3C\,111, which is well known to extend out to kpc-scales \cite{Linfield1984}. This shows that the jet is already well collimated at the resolution limit for VLBA observations at 15\,GHz (from the MOJAVE program\footnote{http://www.physics.purdue.edu/MOJAVE/}, see \cite{Lister2009}) and 43\,GHz (from the Boston University gamma-ray blazar monitoring program\footnote{http://www.bu.edu/blazars/VLBAproject.html}, see \cite{Marscher2011}). The bottom panel depicts the first 86\,GHz GMVA epoch (2007-10-13) closest to the outburst with a prominent bend clearly visible at about 0.5\,mas from the jet base. The angular resolution along the jet is foremost determined by the minor axis of the synthesized beam, leading to an improvement in resolution from 15\,GHz VLBA to 86\,GHz GMVA by about one order of magnitude.
\par
\begin{figure}
\centering
\includegraphics[width=.85\textwidth]{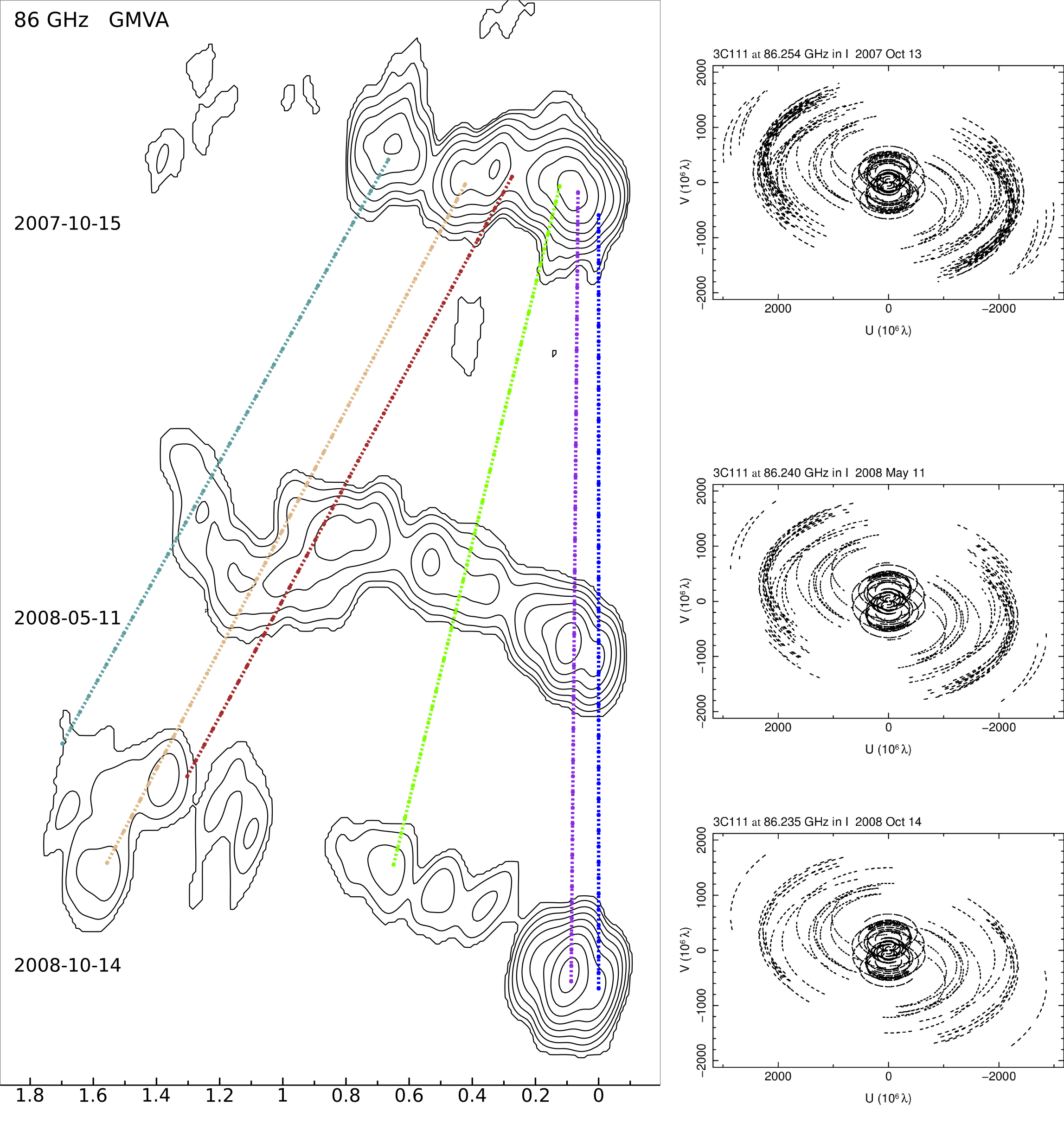}
\caption{\textit{Left panel:} Overview of all three 86\,GHz GMVA epochs of 3C\,111, obtained with hybrid imaging and natural weighting of the visibility data. Top: First epoch (2007-10-13) with a synthesized beam of 0.15$\times$0.045\,mas$^2$ (p.a. $-$11$^\circ$) and a peak flux density of 1.50\,Jy/beam; Middle: Second epoch (2008-05-11) with a synthesized beam of 0.15$\times$0.059\,mas$^2$ (p.a.  $-$16$^\circ$) and a peak flux density of 0.39\,Jy/beam; Bottom: Third epoch (2008-10-14) with a synthesized beam of 0.17$\times$0.067\,mas$^2$ (p.a. $-$18$^\circ$) and a peak flux density of 1.87\,Jy/beam. The lowest contour lines represent 10$\sigma$ (first image) and 5$\sigma$ (second and third image) of the rms noise, increasing logarithmically by factors of two. The images are spaced according to the separation of the observations in time and have been spatially aligned to the jet core, where the emission becomes optically thick. The axis shows distances relative to the core. The coloured lines indicate the evolution of associated features between the epochs based on the preliminary kinematic analysis.\newline
\textit{Right panel:} $(u,v)$-coverage of the GMVA for the images in the left panel. The longest baselines connects the Plateau de Bure Interferometer (France) and Mauna Kea (HI, USA), which gives a $(u,v)$-radius of $\sim$3100\,M$\lambda$ corresponding to $\sim$45\,$\mu$as in angular resolution.}
\label{fig:Overview} 
\end{figure}

All three 86\,GHz images are combined in the left panel of Fig. \ref{fig:Overview}, where the right panel shows the $(u,v)$-coverage of each observation, resulting in unprecedented image fidelity for 3C\,111 at this frequency. On such small scales, the jet evolves dramatically over the period of one year. The bend moves radially away from the core, but remains among the strongest features within the jet in all three observations. The contribution of the inner 0.1\,mas region to the total emission is highest for the third epoch (bottom panel). Interestingly, the jet stream between the core and the bend stays rather straight in the second and third epoch, albeit at a different position angle than in the first epoch. 
\par
A kinematic analysis was performed to study the jet evolution in detail. For this purpose, additional 15 epochs of archival 43\,GHz VLBA data from the Boston University gamma-ray blazar monitoring program between June 2007 and February 2009 were model-fitted for comparison. The complete analysis will be presented elsewhere (R. Schulz et al. in prep.). Preliminary results strongly suggest that the bent feature is associated with the outburst in 2007. At both frequencies, this feature has a position angle between 75$^\circ$ to 80$^\circ$, while the position angles of the rest of the jet remain around 60$^\circ$ to 70$^\circ$. At 86\,GHz, the bend seems to travel with an apparent velocity of about $\sim$3.7\,c (see Fig. \ref{fig:Overview}), which is significantly slower than the highest apparent velocity observed on larger scales at 15 and 43\,GHz \cite{Jorstad2005,Kadler2008}. A complementary long-term study of 3C\,111 is addressing the jet morphology at 15\,GHz based on VLBA data from the MOJAVE program \cite{Grossberger2012}.

%
%

\section{Summary}
We present here images of 3C\,111 based on three successive observations with the Global Millimeter VLBI Array at 86\,GHz following a major outburst in 2007. The images are of unprecedented quality for this object at this frequency, revealing a rapidly evolving bent feature in the jet. Preliminary kinematic results indicate that this bend is related to the outburst in 2007. It moves with a different position angle than other jet features and with a significantly lower apparent velocity than higher-velocity features typically observed on scales beyond 1\,mas in this source.

\acknowledgments
This study makes use of 43 GHz VLBA data from the Boston University gamma-ray blazar monitoring program (http://www.bu.edu/blazars/VLBAproject.html), funded by NASA through the Fermi Guest Investigator Program.
This research has made use of data from the MOJAVE database that is maintained by the MOJAVE team \cite{Lister2009}.
The GMVA data are correlated at the VLBI correlator of the MPIfR,  Bonn, Germany.
The Very Long Baseline Array (VLBA) is an instrument of the National Radio Astronomy Observatory (NRAO). NRAO is a facility of the National Science Foundation, operated by Associated Universities Inc.
ER acknowledges partial support from the Spanish MINECO through grant AYA2009-13036-C02-02, from the Generalitat Valenciana grant PROMETEO-2009-104, and by the COST action MP0905 `Black Holes in a Violent Universe'.

\begin{thebibliography}{99}


\bibitem{Alef1998} W.~Alef, E.~Preuss, K.~I.~Kellermann, et al., \emph{Sub-milliarcsec Structure of 3C 111 at 0.7 and 3.6 CM}, \emph{ASP Conf. Ser.} {\bf 144} \emph{IAU Colloq. 164: Radio Emission from Galactic and Extragalactic Compact Sources} (1998) 129 

\bibitem{Chatterjee2011} R.~Chatterjee, A.~P.~Marscher, S.~G.~Jorstad, et al., \emph{Connection Between the Accretion Disk and Jet in the Radio Galaxy 3C 111}, \emph{ApJ} {\bf 734} (2011) 43 [{\tt arXiv:1104.0663}]

\bibitem{Cohen2007} M.~H.~Cohen, M.~L.~Lister, D.~C.~Homan, et al., \emph{Relativistic Beaming and the Intrinsic Properties of Extragalactic Radio Jets}, \emph{ApJ} {\bf 658} (2007) 232 [{\tt arXiv:astro-ph/0611642}]

\bibitem{Fanaroff1974} B.~L.~Fanaroff, J.~M.~Riley, \emph{The morphology of extragalactic radio sources of high and low luminosity}, \emph{MNRAS} {\bf 167} (1974) 31


\bibitem{Greisen2003} E.~W.~Greisen, \emph{AIPS, the VLA, and the VLBA}, \emph{Information Handling in Astronomy - Historical Vistas} {\bf 285} (2003) 109

\bibitem{Grossberger2012} C.~Grossberger, M.~Kadler, J.~Wilms, et al., \emph{Structural Variability of 3C111 on Parsec Scales}, \emph{AcPol} {\bf 52} (2012) 18, [{\tt arXiv:1110.1197}]

\bibitem{Hartman2008} R.~C.~Hartman, M.~Kadler, J.~Tueller, \emph{Gamma-Ray Emission from the Broad-Line Radio Galaxy 3C 111}, \emph{ApJ} {\bf 688} (2008) 852 [{\tt arXiv:0808.1740}]

\bibitem{Jorstad2005} S.~G.~Jorstad, A.~P.~Marscher, M.~L.~Lister, et al., \emph{Polarimetric Observations of 15 Active Galactic Nuclei at High Frequencies: Jet Kinematics from Bimonthly Monitoring with the Very Long Baseline Array}, \emph{AJ} {\bf 130} (2005) 1418 [{\tt arXiv:astro-ph/0502501v3}]

\bibitem{Kadler2008} M.~Kadler, E.~Ros, M.~Perucho, et al., \emph{The Trails of Superluminal Jet Components in 3C 111}, \emph{ApJ} {\bf 680} (2008) 867 [{\tt arXiv:0801.0617}]

\bibitem{Linfield1984} R.~Linfield, R.~Perley, \emph{3C 111 - A luminous radio galaxy with a highly collimated jet}, \emph{ApJ} {\bf 279} (1984) 60

\bibitem{Lister2009} M.~L.~Lister, H.~D.~Aller, M.~F.~Aller, et al., \emph{MOJAVE: Monitoring of Jets in Active Galactic Nuclei with VLBA Experiments. V. Multi-Epoch VLBA Images}, \emph{AJ} {\bf 137} (2009) 3718 [{\tt arXiv:0812.3947}]

\bibitem{Marscher2011} A.~Marscher, S.~G.~Jorstad, V.~M.~Larionov, et al., \emph{Multi-Waveband Emission Maps of Blazars}, \emph{JApA} {\bf 32} (2011) 233 [{\tt arXiv:1101.0179}]

\bibitem{Marti-Vidal2012} I.~Mart{\'{\i}}-Vidal, T.~P.~Krichbaum, A.~Marscher, et al., \emph{On the calibration of full-polarization 86 GHz global VLBI observations}, \emph{A\&A} {\bf 542} (2012) A107 [{\tt arXiv:1203.1424}]



\bibitem{Shepherd1994} M.~C.~Shepherd, T.~J.~Pearson, G.~B.~Taylor, \emph{DIFMAP: an interactive program for synthesis imaging.}, \emph{BAAS} {\bf 26} (1994) 987

\bibitem{Trippe2011} S.~Trippe, M.~Krips, V.~Pi{\'e}tu, et al., \emph{The long-term millimeter activity of active galactic nuclei}, \emph{A\&A} {\bf 533} (2011) A97 [{\tt arXiv:1107.5456}]

\end{thebibliography}
\end{document}